\documentclass[11pt]{revtex4}
\usepackage[]{epsfig}
\begin{document}


%
%

\title{Negative Pressures in QED  Vacuum in an External Magnetic Field}

\author{H. P\'erez Rojas}

\author{E. Rodr\'{\i}guez Querts}

\affiliation{Grupo de Fisica Teorica, Instituto de Cibern\'etica,
Matem\'atica y F\'{\i}sica,\\
Calle E  No. 309, esq. a 15, Vedado, La Habana 10400, Cuba}



\begin{abstract}
Our aim is to study the electron-positron vacuum pressures in
presence of a strong magnetic field $B$. To that end, we  obtain a
general energy-momentum tensor, depending on external parameters,
which in the zero temperature and zero density limit leads to vacuum
expressions which are approximation-independent. Anisotropic
pressures arise, and in the tree approximation of the magnetic field
case, the pressure along $B$ is positive, whereas perpendicular to
$B$ it is negative. Due to the common axial symmetry, the formal
analogy with the Casimir effect is discussed, for which in addition
to the usual negative pressure perpendicular to the plates, there is
a positive pressure along the plates. The formal correspondence
between the Casimir and black body energy-momentum tensors is
analyzed. The fermion hot vacuum behavior in a magnetic field is
also briefly discussed.
\end{abstract}
\maketitle

\section{Introduction}
Our aim of the present paper is twofold: we are essentially
interested in the study of quantum vacuum (even fermion hot vacuum,
see below) pressures and energies in presence
of a constant external magnetic field $B$. But we have found that
this problem can be considered as a particular case of a more
general problem: the study of the energy-momentum tensor for QED
when there is a symmetry breaking in one (or more) spatial
components, and/or the time component (as it is the case of the
Matsubara temperature formalism).

Concerning the magnetized QED vacuum, we have not found explicit
mention in previous literature to the fact that it exerts
anisotropic pressures, with the remarkable characteristic of having
positive as well as negative values. The anisotropy leads to
\textit{a transfer of momentum from vacuum to real particles or
macroscopic bodies}. This transfer of momentum has been discussed in
the case of a field having electric and magnetic components by
starting from classical grounds \cite{Feigel}. We have found that
the problem of vacuum in a constant homogeneous magnetic field
(which is physically realizable in small regions of QED vacuum in
the more general case of $B$ depending on the spatial coordinates)
has some physical and methodological analogy with Casimir effect
\cite{Casimir},\cite{Bordag}. In both problems appear negative as
well as positive pressures, and they bear the common property of
having axial symmetry determined respectively by the external field
pseudo-vector $B$ and the vector perpendicular to the plates ${\cal
P}_3=2\pi \hbar/d$, $d$ being the distance between the plates. There
are also, however, important physical differences among these
problems, as indicated below. There is otherwise, a complete
correspondence between the energy-momentum tensor in Casimir effect
and in the blackbody radiation problem if coordinates $x_3,x_4$ are
exchanged and the substitution $kT \to \hbar c/2d$ is made. Even
more, there is a full correspondence between the so-called fermion
hot vacuum in a magnetic field and the problem of a magnetic field
perpendicular to the Casimir plates, discussed in Ref.\cite{Cougo},
and all these problems contain as a common property of having broken
the symmetry of one (or more) space-time coordinates.
Correspondingly, some symmetry breaking parameter appears through
non-vanishing field derivatives, which makes  some components of the
four momentum vector $p_{\mu}$ to have discrete values as multiples
of some characteristic parameter arising from the breaking of the
symmetry . In the magnetic field and Casimir cases, the discrete
values appear in some of the spatial components of $p_i$,
$(i=1,2,3)$ and in the temperature case, it is $p_4$ the discrete
term.

\section{The Energy-momentum Tensor}
We discuss in what follows the vacuum energy-momentum tensor by
starting from the quantum relativistic matter tensor in the
temperature formalism, which contains the contribution of vacuum.
Although initial calculations are made in Euclidean variables, where
$x_4$ is taken as some ``imaginary time", vectors and tensors will
be written later by using covariant and contravariant indices.

We will consider the usual QED Lagrangian density $L$ at finite
temperature $T$ and with conserved number of fermions
$N=Tr\gamma_0\int d^3 x \bar \psi({\bf x})\psi({\bf x})$, having
associated chemical potential $\mu$. One can write in an arbitrary
moving frame the density matrix as
\begin{equation}\label{density matrix}
  \rho= e^{-\beta( u_\nu {\cal
P}^\nu-\mu u_\nu J^\nu) }
\end{equation}
 where ${\cal P}^\nu$ is the momentum
four-vector, $J^\nu=N u^\nu$, and $u_\nu$ is the four-velocity of
the medium. From $\rho$, working in the rest system (the spatial
part of the four velocity $u_{\nu}$ vanish), by calling ${\cal
L}=\int d^3 x L$, one gets the effective partition functional as
\begin{equation}
{\cal Z}=C(\beta)\int e^{-\int_0^{\beta}dx_4 \int d^3 x L_{eff}}
{\cal D}_i A_{\mu} {\cal D}\bar\psi {\cal D}\psi \delta (G) Det
{\cal P}.
\end{equation}
Here $A_{\mu}$ is the electromagnetic field, $\bar\psi, \psi$ are
fermion fields, $C(\beta)$ is a normalization constant. The gauge
condition is $G$ and ${\cal P}$ is the (trivial) Fadeev-Popov
determinant \cite{Bernard}, \cite{Kapusta},\cite{Kalashnikov}.  We
have also $L_{eff}= L_{(\partial_{\nu} \to
\partial_{\nu}-\mu\delta_{\nu 4})}$.

The fourth derivative of fermions is shifted in $\mu$: the chemical
potential enter into the density matrix through the vector
$c^{(1)}_{\nu} =\mu u_{\nu}$, and the temperature through
$c^{(2)}_{\nu}=T u_{\nu}$. Then, in the rest system, the field
operators depend on the coordinate ``vectors" $x_\nu=({\bf x},x_4)$,
multiplied by the "momentum" vectors $P_{M\nu}=({\bf p},p_4)$, where
$p_4$ are the Matsubara frequencies, which are
 $2n \pi T$ for bosons and $(2n+1)\pi T$ for
fermions, where $n=0,\pm 1,\pm 2..$.  By taking the quantum
statistical average through the functional integration, indicated
by the symbol $<<..>>$, we obtain the thermodynamical potential
\begin{equation}\label{TDpot}
  \Omega =-\beta ^{-1}\ln <<e^{-\int_0^\beta dx_4 \int
d^3xL_{eff}(x_4, {\bf x)}}>>.
\end{equation}
We observe that the statistical average leads to $<< \int dx_4\int
d^3 x L_{eff}>> \to \Omega $.

The energy-momentum tensor of matter
 plus vacuum will be obtained as a diagonal tensor
(no shearing stresses occur in our approximation) whose spatial
part contains the pressures and the time component is minus the
internal energy density $-U$.
 The total energy-momentum tensor is obtained after quantum averaging as

\begin{equation}\label{total en.mom.tensor}
 {\cal T}^{\mu}_{ \nu }=<<\frac{\partial  L} {\partial a_{i ,\mu
}}a_{i ,\nu} -\delta ^{\mu}_{\nu } L>>,
\end{equation}
where the index $i$ denotes the fields (either fermion or vector
components). One easily gets $<<\int dx_4 \int d^3 x (\partial  L
/\partial \psi_{,4})\psi_{,4}>> =T\partial \Omega/\partial T+\mu
\partial \Omega/\partial \mu$.

By assuming that some derivatives $a_{i,\lambda}$ are non
vanishing quantities, one gets the thermodynamical expression
\begin{equation}
{\cal T}^{i}_{j}=\frac{\partial \Omega}{\partial
a_{i,\lambda}}a_{j,\lambda}-\Omega \delta^i_j,\hspace{1cm}{\cal
T}^{4}_{4}=-(TS+\mu N+\Omega)=-U, \label{tg}
\end{equation}
where ${\cal T}_{ij}$ ($i,j=1,2,3$) are the  pressures and $S=-\partial
\Omega/\partial T$ is the entropy density, $N=-\partial
\Omega/\partial \mu$ is the density of particles and $U$ the
internal energy density.

Relativistic quantum statistical averages in the limit $T \to 0$,
$\mu \to 0$ (see e.g. Fradkin \cite{Fradkin} leads to  the quantum
field averages in vacuum, $<<..>> \to <..>$. The contribution of
observable particles, given by the statistical term $\Omega_{s}(
T, \mu)$ in the expression for the total thermodynamic potential
$\Omega=\Omega_{s}+\Omega_{0}$, vanishes in that limit. The
remaining term leads to the zero point energy of vacuum
$\Omega_{0}$. Thus in the limit $T \to 0$, $\mu \to 0$ and
$<a_{i,\mu}>=0$ (\ref{tg}) leads to the energy-momentum tensor of
vacuum, where ${\cal T}^4_4= -\Omega=-U$ is the vacuum energy
density and the isotropic pressure are
\begin{equation}
{\cal T}^{i}_j=-\Omega \delta^i_j \label{isovac}
\end{equation}
 and we conclude that {\it for the isotropic vacuum, if the
energy density $\Omega>0$, the pressures would be negative} (and
on the opposite, if $\Omega<0$, ${\cal T}^{i}_j>0$)

\section{Vacuum zero point energies}
The solution of the Dirac equation for an electron (positron) in
presence of an external magnetic field $B_j$ for, say, $j=3$,
leads to the energy eigenvalues
\begin{equation}\label{vacen}
\varepsilon_{n}=\sqrt{c^2 p_{3}^{2}+m^{2}c^4+2e\hbar cB n},
\end{equation}
 where $n=0,1,2...$ are the
Landau quantum numbers, $p_{3}$ is the momentum component along
the magnetic field ${\bf B }$ and $m$  is the electron mass.  The
system is degenerate with regard to the coordinates of the orbit´s
center \cite{Johnson}. The spinor wavefunctions and spectrum have
associated a characteristic length is $\lambda_{L}(B)=\sqrt{\hbar
c/2eB}$, and this is valid also for the zero point modes of
vacuum. The electron-positron zero point vacuum energy in an
external electromagnetic field was obtained by  Heisenberg and
Euler \cite{Euler}.  For the case of a pure magnetic field, the
zero point energy density in the tree level approximation
\begin{equation}\label{omega0e}
\Omega_{0e} = \frac{\alpha B^2}{8\pi^2}\int_0^{\infty}e^{-B_c
x/B}F(x)_{HE}\frac{d x}{x},
\end{equation}
 where
 \begin{equation}\label{fe}
 F(x)_{HE}= \left[\frac{coth
x}{x} -\frac{1}{x^2}-\frac{1}{3}\right].
\end{equation}
where $B_c = m_e^2 c^3/e\hbar \sim 4.41 \cdot 10^{13}$ G is the
critical QED magnetic field. The results of the present paper
concern mainly with fields $B \leq B_c$.

In the Casimir effect the motion of virtual photons perpendicular
to the plates is bounded and we have the photon energy eigenvalues
\begin{equation}\label{encas}
\varepsilon_s=c\sqrt{p_1^2+p_2^2+({\cal P}s)^2}.
 \end{equation}
Due to the breaking of the rotational symmetry, the $p_3$
components of the vacuum modes are discrete, $p_3={\cal P}s$ where
$s=0,\pm 1, \pm 2,...$. Here ${\cal P}= 2\pi \hbar/d$, where $d$
is the distance among the plates. This makes the zero point
electromagnetic modes inside the box axially symmetric in momentum
space.  Thus, in both problems there is a quantity characterizing
the symmetry breaking, (and the wavelength in some direction).
These quantities are respectively, the pseudo-vector $B_i$,
determining $\lambda_{L}(B)$, and the basic vector momentum ${\cal
P}_i={\cal P}\delta_{3i}$, perpendicular to the plates (taken
parallel to the $x_1, x_2$ plane). Observe that the term $p_3^2$
in the Casimir effect is proportional to $\hbar^2$, whereas in the
magnetic field case, the term replacing $p_{\perp}^2$ is
proportional to $\hbar$. After taking the sum over Casimir modes
and subtracting the divergent part \cite{Milonni}, one gets a
finite negative term, the well known Casimir energy density
\begin{equation}\label{omegacas}
\Omega_{0C}= -\frac{\pi^2 \hbar c}{720 d^4}=-\frac{c{\cal
P}^{4}}{720 \pi^2\hbar^3 }.
\end{equation}

\section{Vacuum pressures}
\subsection{Magnetized vacuum}
We will consider the vacuum case, when the Lagrangian depends on a
non vanishing field derivative, as it happens when there is an
external field $a_{\mu}=A_{\mu}=B[-x_2, x_1,0,0]/2$ describing a
constant magnetic field (taken along the $3$-rd axis),  which
generates non-vanishing spatial tensor terms through the gauge
invariant electromagnetic field tensor
$<A^{\nu}_{,\mu}-A^{\mu}_{,\nu}>={\cal F^{\mu}_{\nu}}$. From the
previous expression for $\Omega_{0e}(<0)$, this leads to
anisotropic pressure terms of form $P_{0 \perp}={\cal
T}^i_j=-\Omega_{0e} \delta^i_j-{\cal F}^{i}_{k}(\partial
\Omega_{0e}/\partial {\cal F}_{k}^{j})$, or
\begin{equation}
P_{0 \perp}={\cal T}^{1}_{1}={\cal T}^{2}_{2}={\cal T}_{\perp}
=-\Omega_{0e}-B{\cal M}, \label{MF}
\end{equation}
where ${\cal M}=-\partial {\Omega_{0e}}/\partial B$, is the vacuum
magnetization, and $i=1,2$, $j=2,1$. The anisotropy is due  to the
arising of a negative transverse pressure,  generated by an axial
``force": the quantum analog of the Lorentz force, arising when the
magnetic field acts on charged particles having non-zero spin
\cite{Aurora}, and leading to a magnetization of vacuum parallel to
$\textbf{B}$.  The component
\begin{equation}\label{PL}
{\cal T}_{3}^{0e3}=P_{03} = -\Omega_{0e}
\end{equation}
is the pressure along the magnetic field $B$. We
remark that eqs. (\ref{MF}), (\ref{PL}) are approximation-independent.

 One loop
calculations give
\begin{equation}
{\cal M}_{0e}=-2\Omega_{0e}/B- (\alpha
B_c/8\pi^2)\int_0^{\infty}e^{-B_c x/B}F(x)_{HE} d x.
\end{equation}
  It is easy
to verify that ${\cal M}_{0e}>0$. Thus, vacuum shows a
paramagnetic behavior. Concerning the transverse pressures $P_{0e
\perp}=-\Omega_{0e} -B {\cal M}_{0e} $, we get
\begin{equation}\label{pressure}
{\cal T}_{\perp}^{0e}=P_{0e \perp}=\Omega_{0e} + (\alpha B_c
B/8\pi^2)\int_0^{\infty}e^{-B_c x/B}F(x)_{HE}d x.
\end{equation}
Both terms in (\ref{pressure}) are negative, thus, $P_{0e
\perp}<0$. This leads to magnetostrictive effects for any value of
the magnetic field $B$ since QED vacuum is compressed
perpendicularly to $B$, due to the negative pressures, and as the
pressure $P_{03}$ is positive, it is stretched in along $B$. This
could be tested by placing a body (non necessarily metallic)
parallel to it. It would be compressed in the direction
perpendicular to $B$. This is reasonable to expect: the virtual
electrons and positrons are constrained to bound states in the
external field, but moves freely in both directions along the
field. For a wide range of values of $B$, we have that
$\Omega_{0e}<0$ and ${\cal M}_{0e}>0$ holds also if radiative
corrections are included (see below).

As fields currently achieved in laboratories are very small
if compared with the critical field $B_{c}$, in the limit $B<<
B_c$ one can write,
\begin{equation}\label{h}
{\Omega_{0e}}\approx-\frac{\alpha
B^{4}}{360\pi^{2}B_c^2}=-\frac{\pi^{2} \hbar c}{5760 b^{4}},
\end{equation}
where the characteristic parameter is $b(B)= \pi
\lambda_{L}^{2}/\lambda_{C}$. Here
  $\lambda_{L}$ is the magnetic wavelength defined previously
and $\lambda_{C}$ is the Compton wavelength
$\lambda_{C}=\hbar/mc$.  The energy density is then a
  function of
  the field dependent parameter $b(B)$.
we can approximate (\ref{pressure}) as $P_{0e \perp} \approx
3{\Omega_{0e}}<0$. It can be written
\begin{equation}\label{pv}
  P_{0e \perp}\approx
-\frac{\pi^{2} \hbar c}{1920 b^{4}},
\end{equation}
 For small $B$ fields
of order $10-10^3$G, $P_{0e \perp}$ is negligible as compared with
the usual Casimir pressure. But for larger fields, e.g. for $B\sim
10^{5}$ G it becomes larger; one may obtain then pressures up to
$P_{0e \perp} \sim10^{-9} dyn$ $ cm^{-2}$ . For a distance between
plates $d=0.1cm$, it gives $P_{0C} \sim 10^{-14}dyn$ $cm^{-2}$,
(see below) i.e., five orders of magnitude smaller than $P_{0e
\perp}$. This is interesting since a test body (even non metallic)
placed on the magnetized vacuum is stretched along the field and
compressed orthogonal to it. The vacuum stretching effect could be
tested more easily at present times, since magnets producing large
field intensities are being constructed. (for instance, those
being used at CERN for the LHC collider, reaching fields near
$10^5$ G).

The consideration of the next order loop approximation
 would lead to terms having the same field dependence structure,
 for small as well as large fields. These corrections are of
order $\alpha$ times the one loop case (and the same sign),
irrespective of the magnitude of the field. In particular, for
very large fields $B \gg B_c$,
 $\Omega_{0e} \approx -\frac{\alpha B^2}{24\pi^2}\left( \ln\frac{B}{B_c}
-C_{1}\right)$, where  $C_{1}=2.29191...$ is a constant. The
two-loop correction is adapted to our notation from calculations
obtained by Ritus\cite{Ritus}, and it leads in the small field
limit to $\Omega'_{0e} \sim -\frac{ \alpha^2 B^4}{324 \pi B_c^2}$
and for large fields to $\Omega'_{0e} \sim
-\frac{\alpha^{2}B^{2}}{128\pi^{4}}\left( \ln\frac{B}{B_c}
+C_2\right)$, where $C_{2}$ is another  constant. Thus
$\Omega_{0e}$ and $\Omega'_{0e}$ have the same functional
dependence on $B$, being $\Omega'_{0e} \sim \alpha \Omega_{0e}$.
We expect that this behavior is kept for higher order loops, which
does not change essentially our results.

\subsection{Casimir and black body tensors}
In the Casimir effect case, the field derivatives $A_{\mu,3}$ in
momentum space contain the discrete sums in analogy to the
temperature case. Starting from the Casimir energy density, we
have the following pressures
\begin{equation}\label{caspressure1}
 {\cal T}^{3}_{3}=P_{C3}={\cal P}\frac{\partial
\Omega_{0C}}{\partial {\cal P}}-\Omega_{0C} =
3\Omega_{0C}=-\frac{\pi^2 \hbar c}{240 d^4}<0,
\end{equation}
 which is the usual Casimir negative pressure and
\begin{equation}
{\cal T}_{\perp}^C=P_{C\perp}= -\Omega_{0C} =\frac{\pi^2 \hbar
c}{720 d^4}>0 \label{Milon3}
\end{equation}
which is a  positive pressure acting parallel to the plates in the
region inside them. This is a second Casimir force. (This is not the
so-called lateral Casimir force described in
Ref.\cite{Mostepanenko}).

We remind that for black body radiation in equilibrium
\cite{Landau1} it is ${\cal T}_{bj}^i=-\Omega_b \delta^{i}_{j}$,
where $i,j=1,2,3$, and ${\cal T}_{b4}^4=-U_b=3\Omega_b=-\pi^2
T^4/15\hbar^3 c^3$. One must remark that the usual Casimir
pressure corresponds to minus the energy of the black body
radiation at $T=T_{Cas}$, e.g., $P_{C3}={\cal T}^{C3}_{3} \to
{\cal T}^4_{b4}=-U_b$  and the Casimir energy corresponds to minus
the black body pressure ${\cal T}^{C4}_{4}= -\Omega_{0C}\to
-\Omega_b$, that is, both tensors are similar under exchange of
their ${\cal T}^3_3, {\cal T}^4_4$ components.

\subsection{Magnetized fermion hot vacuum} We use the name ``hot vacuum" in statistical QED to describe an
electrically neutral electron-positron gas at temperature $T$ (for
which obviously the chemical potential $\mu= 0$) in thermal
equilibrium with black body radiation \cite{Shabad}. If $T \to 0 $,
the density of these particles decrease to zero and one is left with
the usual quantum vacuum limit. Hot vacuum conditions may occur for
instance around a neutron star if thermal radiation coexists with
$e^{\pm}$ pairs due to $\gamma$ radiation decay.  If it is placed in
an external magnetic field, its fermionic thermodynamic potential
$\Omega_{fT}$ describes a neutral gas of magnetized electrons and
positrons, which grows with $T$. In the limit $T \to 0$ one gets in
the quantum relativistic temperature formalism \cite{Shabad} the
usual vacuum Euler-Heisenberg energy density term. In studying the
magnetic properties in the first loop approximation, there is no
contribution from the photon gas in equilibrium with the
electron-positron system, since in that approximation, the photon
interaction with the magnetic field is not included.However, the
total hot vacuum thermodynamic potential must take into account the
background electromagnetic blackbody radiation
$\Omega_{HV}=\Omega_{fHV}+\Omega_{bHV}$,\hspace{.5cm}
$\Omega_{bHV}=\frac{2}{(2\pi^{3})\beta} \int \ln
(1-e^{-\sqrt{p^2}\beta})d^{3}p=-\frac{\pi^{2}}{45\beta^{4}}$. The
thermodynamic potential is then  $\Omega_{fHV}=\Omega_{0
e}+\Omega_{fT}$, where $\Omega_{fT}$ is the temperature fermion
sector (in which $\beta=\frac{1}{kT}$)
\begin{equation}\label{hotvac}
   \Omega_{fT}=\frac{eB}{4\pi^{2}(\hbar
  c)^{2}}\sum_{n=1}^{\infty}(-1)^{n}\int_{0}^{\infty}\frac{dt}{t^{2}}
  e^{-m^{2}c^{4}t-\frac{\beta^{2}n^{2}}{4t}}\coth (eB\hbar ct).
\end{equation}
Notice that this term is analog to the Casimir -Euler-Heisenberg
contribution to the effective Lagrangian obtained by Cougo-Pinto
et.al. in Ref.\cite{Cougo} for the Casimir effect in presence of a
magnetic field, if the transformation $kT \to \hbar c/2d$ is
performed ($d$ is the separation between plates).
\begin{figure}[pb]
\centerline{\psfig{file=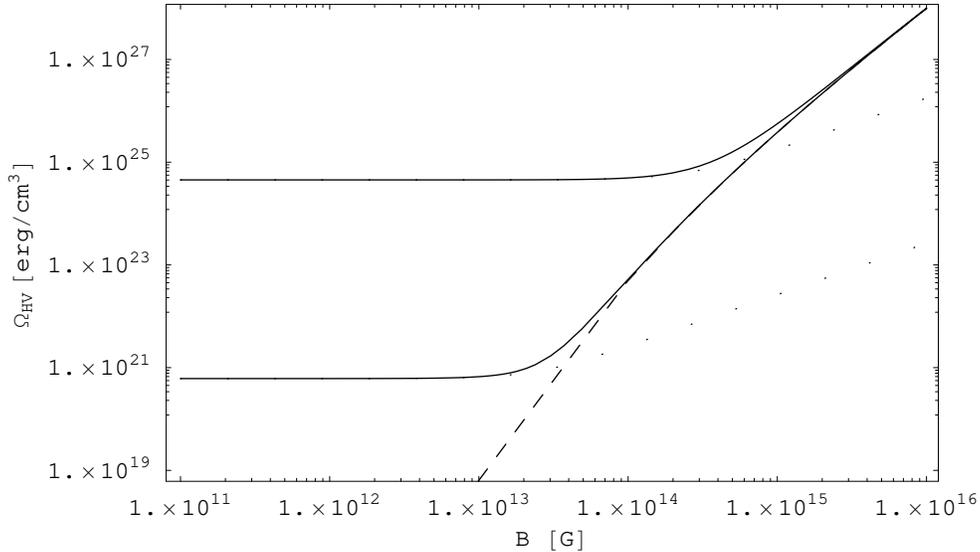,width=13cm}} \vspace*{8pt}
\caption{Fermion hot vacuum thermodynamic potential dependence on
the magnetic field $B$ for $T=.2Tc=1.186\cdot10^{9}K$ (lower
curve), $T=Tc=\frac{mc^{2}}{k}=5.930\cdot10^{9}K$ (upper curve).
Dashed line corresponds to the pure vacuum contribution $\Omega_{0
e}$ and with dotted line we represents $\Omega_{fT}$. It is easy
to see that for $eB\hbar c\gg (kT)^{2}$, $\Omega_{fHV}\approx
\Omega_{0 e}$, and for $eB\hbar c \lesssim (kT)^{2}$,
$\Omega_{fHV}\approx \Omega_{fT}$. \label{fig}}
\end{figure}

From the expression for the thermodynamic potential, we get the
pressure in the direction along the magnetic field $P_{fHV
3}=-\Omega_{fHV}>0$, and the hot vacuum magnetization ${\cal
M}_{fHV }=-\partial \Omega_{fHV}/ \partial B= {\cal M}_{0e}+{\cal
M}_{fT}$, with
\begin{equation}\label{HVmagn}
  {\cal M}_{fT}=-\frac{\Omega_{fT}}{B}+\frac{(\alpha B)}{4\pi^{2}\hbar
  c}\sum_{n=1}^{\infty}(-1)^{n}\int_{0}^{\infty}\frac{dt}{t}
  e^{-m^{2}c^{4}t-\frac{\beta^{2}n^{2}}{4t}}\sinh^{-2} (eB\hbar
  ct).
  \end{equation}
   The vacuum magnetic response conduces to the achievement of anisotropic
   pressures: the perpendicular
   pressure is $P_{HV \perp}=-\Omega_{HV}-B{\cal M_{HV}}=P_{0 e \perp}+P_{T
\perp}$, where
\begin{equation}\label{HVtranspr}
  P_{fT \perp}=-\frac{(eB)^{2}}{4\pi^{2}\hbar
  c}\sum_{n=1}^{\infty}(-1)^{n}\int_{0}^{\infty}\frac{dt}{t}
  e^{-m^{2}c^{4}t-\frac{\beta^{2}n^{2}}{4t}}\sinh^{-2} (eB\hbar ct).
\end{equation}
The total hot vacuum energy is given by
\begin{equation}
U_{fHV}=T\frac{\partial \Omega_{fHV}}{\partial T}-\Omega_{fHV},
\end{equation}
which, under the exchange $kT \to \hbar c/2d$, leads to the
Casimir electron-positron pressure under the action of the
magnetic field plus the usual virtual photon pressure. Then the
total pressure along the magnetic field is
\begin{equation}
P_{f3}=U_{fHV}(kT = \hbar c/2d).
\end{equation}

It is easy to check that for strong fields (or small temperatures),i.e., when  $eB\hbar c
  \gg (kT)^{2}$ holds,
  the temperature effects can be neglected since $\Omega_{0 e}\gg
  \Omega_{fT}$ (see Fig.~\ref{fig}), and the system behaves very similar to the pure magnetized
  vacuum. This result might be of interest for instance in the study of
  neutron stars,  if we assume its magnetosphere as composed by an electron-positron gas
  in chemical equilibrium with decaying $\gamma$ quanta.
  For magnetic field intensities satisfying the previous
  inequality, the magnetic properties of neutron star magnetosphere pressures behave
  like those of the magnetized quantum vacuum case.
  When $eB\hbar c \ll (kT)^{2}$, we have $\Omega_{fHV}\approx
  \Omega_{fT}$. Then the transverse pressure becomes positive and tends to
  equalize the parallel pressure: the  behavior slightly deviates from the
  isotropic black-body case. The reader can check that a similar analysis is valid for the
  Casimir effect in a magnetic field, for $eB\hbar c \gg (\hbar c/2d)^{2}$
  and $eB \hbar c \ll (\hbar c/2d)^{2}$,
  respectively.

\section{Conclusions}
The effect of a constant homogeneous magnetic field modify QED
vacuum zero point energy and magnetize the vacuum. We have reasons
to expect that higher order corrections does not changes the
essence of this behavior. A positive
pressure is exerted in the direction parallel to $B$, while
negative perpendicular to $B$ pressures appear. This means that vacuum
shrinks perpendicular to the magnetic field and freely flows
parallel to it. This could be tested by placing a body (non
necessarily metallic) in the magnetized vacuum. It would be compressed in
the direction perpendicular to $B$. This is reasonable to expect:
the virtual electrons and positrons are constrained to bound
states in the external field, but flows freely in both directions
along the field. That motion of virtual particles can be
interpreted as similar to the real electrons and positrons,
describing "orbits" having a characteristic radius of order
$\lambda =\sqrt{\hbar c/eB}$ in the plane orthogonal to $B$, but
the system is degenerate with regard to the position of the center
of the orbit. One must remark that our results can be applied to
the case in which the magnetic field is inhomogeneous, if the
region in study is divided in volumes small enough such that the
field can be considered approximately homogeneous in each of them.
Then it is easy to conclude that the transverse pressure will
increase towards the region where the field is larger. For
instance, if the field has axial symmetry, having its maximum
value on that axis, and decreasing as we move out from it, the
resultant negative pressures will be directed to that axis of
symmetry.

 The arising of negative pressures in the magnetic field case bears some analogy to the
 Casimir effect, and correspondingly, in addition to the well known negative
pressure, a second positive Casimir pressure acting parallel to
the plates is obtained. The combined effect of these two pressures
leads to a flow of vacuum parallel to plates which would mean a
transfer of momentum to test particles located in the cavity.

 After
studying  temperature effects on the properties of magnetized QED
vacuum, we conclude that for $eB\hbar c \gg T^{2}$ hot vacuum
shows an anisotropic behavior. The transverse  pressure becomes
negative, due to the effect of vacuum magnetization.  For $eB\hbar
c \ll T^{2}$ hot vacuum behaves very much like usual isotropic  black body
radiation.

 In all the three studied  cases the
anisotropy is achieved due to the symmetry breaking in one or more
space-time directions.
\section{Acknowledgements}
Both authors thank H. Mosquera Cuesta and A.E. Shabad for
discussions. H.P.R. is especially indebted to G. Altarelli, M.
Chaichian, C. Montonen,  D. Schwartz and R. Sussmann for
enlighting comments. He also wants to thank CERN, where part of
this paper was written, for hospitality. .

\end{document}